\documentstyle[preprint,aps]{revtex}
\draft
\begin{document}
\title{Resonant tunneling through a macroscopic charge state 
in a superconducting SET transistor} 
\author{D.V. Averin}
\address{Department of Physics, SUNY at Stony Brook, Stony Brook, 
NY 11794, USA}
\author{A.N. Korotkov}
\address{Institute of Nuclear Physics, Moscow State University, 
119899 GSP Moscow, Russia} 
\author{A.J. Manninen and J.P. Pekola}
\address{Department of Physics, University of Jyv\"{a}skyl\"{a}, 
P.O. Box 35, 40351 Jyv\"{a}skyl\"{a}, Finland}
\maketitle

\begin{abstract}
We predict theoretically and observe in experiment that the 
differential conductance of a superconducting SET transistor 
exhibits a peak which is a complete analogue 
in a {\em macroscopic} system of a standard resonant tunneling 
peak associated with tunneling through a single quantum state. 
In particular, in a symmetric transistor, the peak height is 
universal and equal to $e^2/2\pi \hbar$. Away from the resonance  
we clearly observe the co-tunneling current which in contrast to 
the normal-metal transistor varies linearly with the bias voltage.    
\end{abstract}

\vspace{3ex} 

\pacs{PACS numbers: 73.40.Gk, 74.50.+r}
\narrowtext

Charging effects in systems of small Josephson junctions are quite 
well understood by now -- see, e.g., \cite{b1,b2}. Interest, 
however, has been focused mostly on the interplay between the 
charging effects and Cooper pair transport, which can be described 
in generic terms as the quantum dynamics of the Josephson phase 
difference. The aim of this work is to study the quasiparticle 
transport in a superconducting SET transistor -- a system of two 
junctions connected in series (see inset in Fig.\ 1). We show that 
the BCS singularity in the density of states of superconducting  
electrodes of the junctions brings about several interesting new 
features of quasiparticle transport. Most notably, in the 
vicinity of the threshold voltage $V_t$ for classical tunneling 
the quasiparticle transport is identical to resonant tunneling 
through a single macroscopic quantum state of the transistor. 

In this work, we study the low voltage regime, $V<V_t$, where 
the quasiparticles do not have enough energy to enter the central 
electrode of the transistor and can traverse it only by quantum 
tunneling through the energy barrier created by the charging energy 
of the central electrode. The effects of the superconducting 
density of states in the classical sequential tunneling were 
discussed recently in Ref.\  \cite{b*}. 
The dominant contribution to the current $I$ in the regime of 
quantum tunneling comes from the so-called inelastic co-tunneling, 
the process in which two different electrons tunnel simultaneously 
in the two junctions of the transistor, and can be written 
\cite{b3} as: 
\[ I=e(\Gamma(V) -\Gamma(-V))\, , \] 
\begin{eqnarray} 
\Gamma(V)= \frac{8\pi}{\hbar} \int \Pi_{j=1}^{4} \left[ d\epsilon_j 
N_j(\epsilon_j) (1-f(\epsilon_j)) \right] \delta(eV- \sum_{j=1}^{4} 
\epsilon_j)  \nonumber \\ 
|T_1T_2|^2 |\frac{1}{E_1 +\epsilon_1 +\epsilon_2 -2\Delta} +   
\frac{1}{E_2 +\epsilon_3 +\epsilon_4 -2\Delta} |^2 \, .  
\label{1} \end{eqnarray} 
Here $\epsilon_1,\, \epsilon_2$ and $\epsilon_3,\, \epsilon_4$ 
are the energies of the states between which electrons are 
transferred in the first and second junction, respectively, 
$T_{1,2}$ and $N_j(\epsilon_j)$ are the corresponding tunneling 
amplitudes and densities of states in the electrodes, $f(\epsilon)$ 
is the Fermi distribution function, $V$ is the bias voltage, and  
we have assumed for simplicity that all electrodes have the 
same energy gap $\Delta$. 

We restrict our attention to the case of low temperatures, $T\ll 
\Delta$, when the non-vanishing quasiparticle current exists 
only at large voltages, $V>4\Delta/e$, sufficient for the creation 
of quasiparticles in the two junctions. In this voltage range 
the energies $E_{1,2}$ of the intermediate charge states in 
eq.\ (\ref{1}) are: 
\begin{equation} 
E_1=E_C -\lambda (eV-4\Delta) - \frac{eQ_0}{C_{\Sigma}}\, , \;\;\;\;   
E_2=E_C -(1-\lambda) (eV-4\Delta) + \frac{eQ_0}{C_{\Sigma}}\, , 
\label{2} \end{equation} 
where $E_C =e^2/2C_{\Sigma}$ with $C_{\Sigma}= C_1+C_2+C_g$ 
denoting the total capacitance of the central electrode of the 
transistor, $\lambda =(C_2+C_g)/C_{\Sigma}$ gives the 
fraction of the bias voltage that drops across the first junction, 
and $Q_0 = e\times \{ V_gC_g/e+ \Delta(2\lambda -1)/E_C \}$ with  
$\{x\} \equiv x- [x+1/2]$ can be interpreted as the charge induced 
by the gate voltage $V_g$ into the central electrode.       

Integrating over $\epsilon_1,\, \epsilon_2$ at fixed $\epsilon_1+ 
\epsilon_2$ and similarly over $\epsilon_3,\, \epsilon_4$ we can 
express the co-tunneling rate (\ref{1}) in terms of the ``seed'' 
$I-V$ characteristics $I_j(U)$, $j=1,2$, of the two junctions at a 
fixed voltage $U$ across a single junction and no charging effects: 
\[ \Gamma(V)= \frac{\hbar}{2\pi e^2} \int d\epsilon \frac{I_1 
(\epsilon/e)}{1-\exp (-\epsilon/T)} \frac{I_2(V-\epsilon/e)}{1- 
\exp (-(eV-\epsilon)/T)} |M|^2 \, , \] 
\begin{equation} 
M = \frac{1}{E_1 +\epsilon -2\Delta} +\frac{1}{E_2 +eV- 
\epsilon -2\Delta} \, . 
\label{3} \end{equation} 

>From eq.\ (\ref{3}) we see directly that a jump of the quasiparticle 
current $I_j(U)$ at $U=2\Delta/e$ in superconducting junctions 
changes the voltage dependence of the co-tunneling current for 
$V$ close to $4\Delta/e$ from cubic ($\Gamma(V) \propto V^3$ for 
a normal-metal transistor \cite{b3}) to linear. Indeed, for $T\ll 
\Delta$ we can approximate $I(U)$ near the threshold  $U=2\Delta/e$ 
as (see, e.g., \cite{b4}):  
\begin{equation} 
I(U) = I_j \Theta(U-2\Delta/e)\, , \;\;\;\; I_j = \frac{\pi \Delta 
}{2e R_j} \, ,  
\label{4} \end{equation} 
where $R_j$ is the normal-state tunnel resistance of the $j$th 
junction. Equations (\ref{3}) and (\ref{4}) give for low 
temperatures and $eV-4\Delta \ll \Delta,\, E_C$:    
\begin{equation} 
I(V) = e\Gamma(V) = \frac{\hbar I_1I_2}{2\pi} (\frac{1}{E_1}+ 
\frac{1}{E_2})^2 (V-\frac{4\Delta}{e}) \, .
\label{5} \end{equation} 

When the bias voltage approaches the threshold $V_t$ of 
classical sequential tunneling, where one of the energy 
barriers $E_j$ vanishes, the co-tunneling current grows and crosses 
over into the current carried by sequential tunneling, in which 
quasiparticles traverse the transistor by two independent jumps 
across the two junctions. It is known that the energy width of the 
crossover region between the co-tunneling and sequential tunneling 
is determined by the lifetime broadening of the intermediate charge 
states $E_{1,2}$ \cite{b5,b6,b7,b8}. If the gate voltage is not 
close to the special point $Q_0=(1/2-\lambda)e$ where $E_{1,2}$ 
vanish simultaneously (the situation that corresponds to the 
maximum threshold voltage $V_t=(4\Delta+2E_C)/e$), then the current 
through one intermediate state, for instance $E_1\equiv E$, 
dominates near the tunneling threshold. The current in the 
transition region can be described in this situation by simply 
adding the lifetime broadening $\gamma$ of the intermediate state 
in eq.\ (\ref{3}) for the co-tunneling rate \cite{b9,b10}:
\begin{equation} 
M = \frac{1}{E +\epsilon -2\Delta+i\gamma} \,, \;\;\; 
\gamma= \frac{\hbar}{2e} [I_1(\frac{\epsilon}{e}) \mbox{coth}  
(\frac{\epsilon}{2T})+I_2(\frac{eV-\epsilon}{e}) \mbox{coth}
(\frac{eV-\epsilon}{2T})] \, . 
\label{6} \end{equation} 
(This simple approach neglects only the renormalization of $E$ 
and $\gamma$ significant at temperatures exponentially small on 
the scale of $E_C$ \cite{b10}.) 

Combining eqs.\ (\ref{2}), (\ref{3}), (\ref{4}), and (\ref{6}) 
we can calculate the differential conductance of the transistor at 
low temperatures:  
\begin{equation} 
G= \frac{dI}{dV} = \frac{\hbar I_1I_2}{2\pi} \left[ \frac{\lambda}
{(E_0 -\lambda (eV-4\Delta))^2 + \delta^2} + \frac{1-\lambda}{(E_0+
(1-\lambda) (eV-4\Delta))^2 + \delta^2}\right] \, ,
\label{7} \end{equation} 
where $E_0=(e/2-|Q_0|)e/C_{\Sigma}$ is the Coulomb energy barrier 
at $V=4\Delta/e$, 
and $\delta=\hbar(I_1+I_2)/2e$ is the energy width of the charge 
state due to tunneling. If we use the second equation in the 
expression (\ref{4}) we see that $\delta=\pi \hbar (R_1^{-1}+ 
R_2^{-1})\Delta /4e^2$. Since the ideology of co-tunneling is 
applicable only to junctions with small tunnel conductance, $R^{-1} 
\ll e^2/h$, this means that the width of the charge state 
is small, $\delta \ll \Delta$, and eq. (\ref{7}) describes  
the narrow conductance peak located at the threshold  $V_t$ of 
classical tunneling ($eV_t=4\Delta+ E_0/\lambda$). This peak 
corresponds to the 
rapid current rise from almost zero to $I_1I_2/(I_1+I_2)$ at 
$V=V_t$. The maximum conductance is achieved when $E_0 =0$ (i.e.,  
when the tunneling threshold reaches minimum) and $V=V_t=4\Delta/e$: 
\begin{equation} 
G= \frac{dI}{dV} = \frac{e^2}{2\pi \hbar} \frac{4I_1I_2}{(I_1
+I_2)^2} \, . 
\label{8} \end{equation} 
Equation (\ref{8}) shows that in a symmetric transistor, where 
$I_1=I_2$, the differential conductance reaches the absolute 
maximum $e^2/2\pi \hbar$ which is independent of $\Delta, \, E_C$, 
or the junction resistance $R$. This universality is similar to 
that of the resonant tunneling through a single microscopic quantum 
state, and is quite remarkable in view of the fact that in the 
present context the quantum state is the {\em macroscopic}  
charge state of the central electrode of the transistor.  

If the energy barrier $E_0$ is large on the scale of the width 
$\delta$ of the charge state, $\delta$ starts to increase with 
increasing $E_0$, i.e. increasing threshold voltage $V_t$. The 
conductance peak can be described analytically in this regime by 
retaining only the first, resonant, 
term in eq.\ (\ref{7}), and taking into account that 
the peak width $\delta$ depends then on its position 
$V_t$ through the dependence on $V_t$ of the contribution of 
the current $I_2$ through the second junction to $\delta$: 
$I_2=I_2(V_t-2\Delta/e)$.  

The shape of the conductance peak in a symmetric transistor 
(with $R_1=R_2$, and $\lambda =1/2$) calculated numerically from 
the eqs.\ (\ref{2}), (\ref{3}), and (\ref{6}) without 
the approximation (\ref{4}) or restrictions on $E_0$ is shown 
in Fig.\ 1. We see that this, more accurate, 
calculation preserves all the qualitative features of 
the simple analytical expression (\ref{7}): maximum conductance is 
$e^2/2\pi \hbar$ when $E_0=0$ and decreases to approximately half 
this value at nonzero $E_0$.

For the results discussed above to be valid, the lifetime 
broadening of the resonant charge state should not only be much 
smaller than the superconducting gap $\Delta$, but also much 
smaller than the typical energy distance (on the order of $E_C$) 
to the excited charge states of the central electrode of the 
transistor. The condition for this is:
\begin{equation}    
\alpha \equiv \frac{\Delta}{E_C} \frac{\pi \hbar}{e^2} 
(R_1^{-1}+ R_2^{-1}) \ll 1 \, . 
\label{9} \end{equation} 
It is important that this condition can be violated not only 
when the normal-state junction conductances are large, but also 
when the energy gap $\Delta$ is large in comparison to the 
charging energy $E_C$. If it is indeed violated, the charging 
effects are eventually washed out by quantum fluctuations and 
the current rise at $V=4\Delta/e$ becomes infinitely sharp 
(provided that the singularity of the density of states at the 
energy gap $\Delta$ is not smeared out by some internal mechanism 
like spin-flip scattering).  

To test these predictions experimentally we fabricated and measured 
four superconducting SET transistors with differing parameters. The 
transistors were fabricated by electron beam lithography on oxidised 
silicon by the standard shadow evaporation technique using aluminum 
electrodes and aluminum oxide junction barriers. The geometry of 
each of the transistors was such that the length of the central 
island was 1 $\mu$m, its width was 80 -- 120 nm, and the overlap at 
the two ends of the island with the external electrodes was nominally 
70 nm. The gate electrode was simply a 100 nm wide finger, pointing 
orthogonally to the center of the island at a distance of about 
0.5 $\mu$m.The gate capacitance was about 0.02 fF. 

Tunnel resistance $R$ of the transistor junctions was measured from 
the large-voltage asymptote of the $I-V$ characteristic 
of the transistor assuming equal resistances of the two junctions. 
Although we did not carry out any systematic study of how symmetric 
the transistors were, we checked from the gate voltage dependence 
of the threshold voltage $V_t$ that sample 1 had equal parameters 
to within 30\%, and we do not expect the other 
transistors to be worse in this respect since their dimensions were 
larger than in sample 1. The charging energy $E_C$ was measured as 
a half of the amplitude of the $V_t$ modulation by the gate voltage, 
and $\Delta$ can be obtained from the onset of the current at 
$4\Delta/e$. All these parameters of the four samples are shown in 
Table 1, together with the combined parameter $\alpha$ defined in 
eq. (\ref{9}) as a small parameter of the present theory. 

Measured $I-V$ characteristics and traces of the differential 
conductance of sample 1 as a function of the bias voltage $V$ 
are shown in Fig.\ 2 for several values 
of the gate voltage. The curves agree qualitatively with the 
predictions of the theory described above. The differential 
conductance has a narrow peak of the roughly correct width at the 
threshold of classical tunneling. The height of the peak away from 
the resonance is slightly below one-half of $e^2/2\pi \hbar$. The 
main discrepancy between the experimental results (Fig.\ 2b) and the 
simple model calculations (Fig.\ 1) is that at resonance the 
conductance does not reach the ideal maximum value $e^2/2\pi \hbar$ 
but rather is about one half of this value. Although the asymmetry 
of junction resistances contributes according to eq.\ (\ref{8}) 
to suppression of the resonance conductance, the actual asymmetry 
of our transistors was too small to account for the observed magnitude 
of this suppression. This discrepancy can be qualitatively explained 
by the fact that 
in close similarity to regular resonant tunneling, the resonant 
tunneling through the macroscopic charge state discussed in this 
work is very sensitive to all sources of inelastic scattering. 
For instance, we checked numerically that weak spin-flip scattering 
with the rate $\tau_s^{-1}=0.01 \Delta/\hbar$ is sufficient to 
suppress the resonant conductance peak to the level found in the 
experiment. The model with spin-flip scattering, however, 
did not reproduce correctly the full shape of the observed 
conductance curves and we think at the moment that in our 
transistor the resonance is suppressed by a combination 
of several inelastic scattering mechanisms including fluctuations 
of the bias and gate voltage (associated with the finite impedance 
of the voltage leads), and inelastic tunneling through the tunnel 
barriers. We could not characterise quantitatively all these 
sources of inelastic scattering, and therefore did not attempt to 
find a theoretical fit to the curves in Fig.\ 2.  
 
The results of measurements for all four samples are 
summarized in the right-hand-part of Table 1, which shows two  
characteristic values of the differential conductance in units 
of $e^2/2\pi\hbar$: (i) 
$G_{0,exp}$, the conductance at bias voltage just above 
$4\Delta/e$ and at gate voltage that corresponds to the maximum 
threshold voltage $V_t$, and (ii) $G_{1,exp}$, the peak conductance  
at resonance (when $V_t$ reaches minimum). Variation 
of the peak conductance $G_1$ with the tunnel resistance $R$ 
and charging energy $E_C$ described by Table 1 confirms that when 
the relative width of the charge states of the transistor 
(characterised by the parameter $\alpha$ of eq.\ (\ref{9})) becomes 
considerable, $G_1$ increases gradually beyond $e^2/2\pi\hbar$. 
At large $\alpha$, when $G_1$ is much larger than $e^2/2\pi\hbar$,  
the charging effects are completely washed out by the quantum 
fluctuations of charge on the central electrode of the 
transistor and the differential conductance becomes 
insensitive to the gate voltage. This case is approached by 
sample 4 with the largest $\alpha$ in which $V_t$ is practically 
independent of the gate voltage, and we could assign only one 
value of the characteristic conductance to this sample.  

When $\alpha$ is small and the charging effects are 
well-pronounced, the threshold conductance $G_0$ originates  
only from the process of co-tunneling, and is much smaller than 
the peak conductance $G_1$. It can be calculated from eq. 
(\ref{5}) which predicts that at $V=4\Delta/e$ the co-tunneling 
conductance of the superconducting SET transistor should 
increase abruptly to a finite, voltage-independent level which also  
does not depend on temperature at low temperatures. This behavior is 
indeed found in our three samples with larger tunnel resistances. 
Figure 3 shows for example the data for sample \# 2. At gate 
voltages which correspond to the thresholds $V_t$ close to maximum 
we see the kink in the $I-V$ curves and the step in the $dI/dV$ at 
the onset of the quasiparticle current at $V=4\Delta/e$. 
(For other values of the gate voltage small current peaks due to 
Cooper pair tunneling that are visible in Figs.\ 2a and 3, overlap 
with the onset of quasiparticle current and do not allow to identify 
the conductance jump.)  The data shown in Figs.\ 2 and 3 were taken 
at temperature of about 100 mK. We checked that the jump in the 
quasiparticle conductance is practically temperature-independent 
for temperatures up to 0.4 K. 

Table 1 contains a comparison between the observed co-tunneling 
conductance  $G_{0,exp}$ and $G_{0,theory}$ calculated from 
eq.\ (\ref{5}) under the assumption of a symmetric transistor.
Taking into  account that any asymmetry of the junction 
tunnel resistances increases $G_0$ we can say that the agreement 
between $G_{0,exp}$ and $G_{0,theory}$ is reasonable. 

In summary, we proposed theoretically and confirmed in experiment 
that the quasiparticle transport in a superconducting SET 
transistor in the vicinity of the tunneling threshold can be 
described as resonant tunneling through a macroscopic charge state 
of the central electrode of the transistor. The maximal differential 
conductance associated with this process is $e^2/2\pi \hbar$, while 
the width of the resonance is determined by the lifetime broadening 
of the charge states of the transistor. For gate voltages away from 
the resonance we observed very clearly the co-tunneling current 
which exhibits linear (in contrast to cubic of the normal-metal 
case) dependence on the bias voltage. 

We gratefully acknowledge financial support of the Academy of 
Finland and US AFOSR, and Mikko Leivo for help with the experiments.

\figure{Calculated bias-voltage dependence of the differential 
conductance of a symmetric superconducing SET transistor with 
junction resistance $R=20 \hbar/e^2$. The curves are plotted for 
several values of the gate voltage, i.e., the charge $Q_0$ induced 
on the central electrode of the transistor, that correspond to 
several charging energy barriers $E_0$ for tunneling: 
$E_0/(\hbar \Delta/R e^2) =0;\, 1;\, 3;\, 6;\, 10$. The induced 
charge $Q_0$ can not be close to 0. The inset shows the equivalent 
circuit of the SET transistor.  \label{f1}}

\figure{Measured (a) $I-V$ characteristics, and (b) bias-voltage 
dependence of the differential conductance of sample 1 for several 
gate voltages. The traces shown with thick lines in (b) correspond 
to the $I-V$ curves presented in (a). For clarity, the features 
due to the current peaks associated with the Cooper-pair tunneling 
that are visible in (a) have been omitted in (b). For discussion 
see text. \label{f2}}

\figure{ Measured $I-V$ characteristics of sample 2 for 
several gate voltages. The inset shows the differential conductance 
in the vicinity of the gap edge $V=4\Delta/e$ for gate voltages 
which correspond to the two largest tunneling thresholds. The 
conductance jump at $V=4\Delta/e$ is due to the co-tunneling.   
\label{f3}}

\begin{table} 
\begin{tabular}{|c|c|c|c|c|c|c|c|} \hline 
sample & $R($k$\Omega$) & $E_C$ (meV) & $\Delta$ (meV) & 
\hspace*{1.0em} $\alpha $ \hspace*{1.0em} & $G_{0,exp}$ &  $G_{0,theory}$ & 
$G_{1,exp}$ \\ \hline 

1 & 206 & 0.35 & 0.22 & 0.08 & 3.1 $\cdot 10^{-3}$ &  
1.6 $\cdot 10^{-3}$ & .5 \\ \hline  

2 & 152 & 0.15 & 0.21 & 0.24 & 0.032 & 0.014 & 0.9 \\ \hline 

3 & 65 & 0.15 & 0.20 & 0.55 & 0.096 & 0.086 & 1.6 \\ \hline 

4 & 52 & 0.08 & 0.23 & 1.44 & -- & -- & 4 .0  \\ \hline  
\end{tabular} 
\caption{Parameters of the four studied SET transistors. Conductances 
in the last three columns are shown in units of $e^2/2\pi\hbar$. }  
\end{table} 

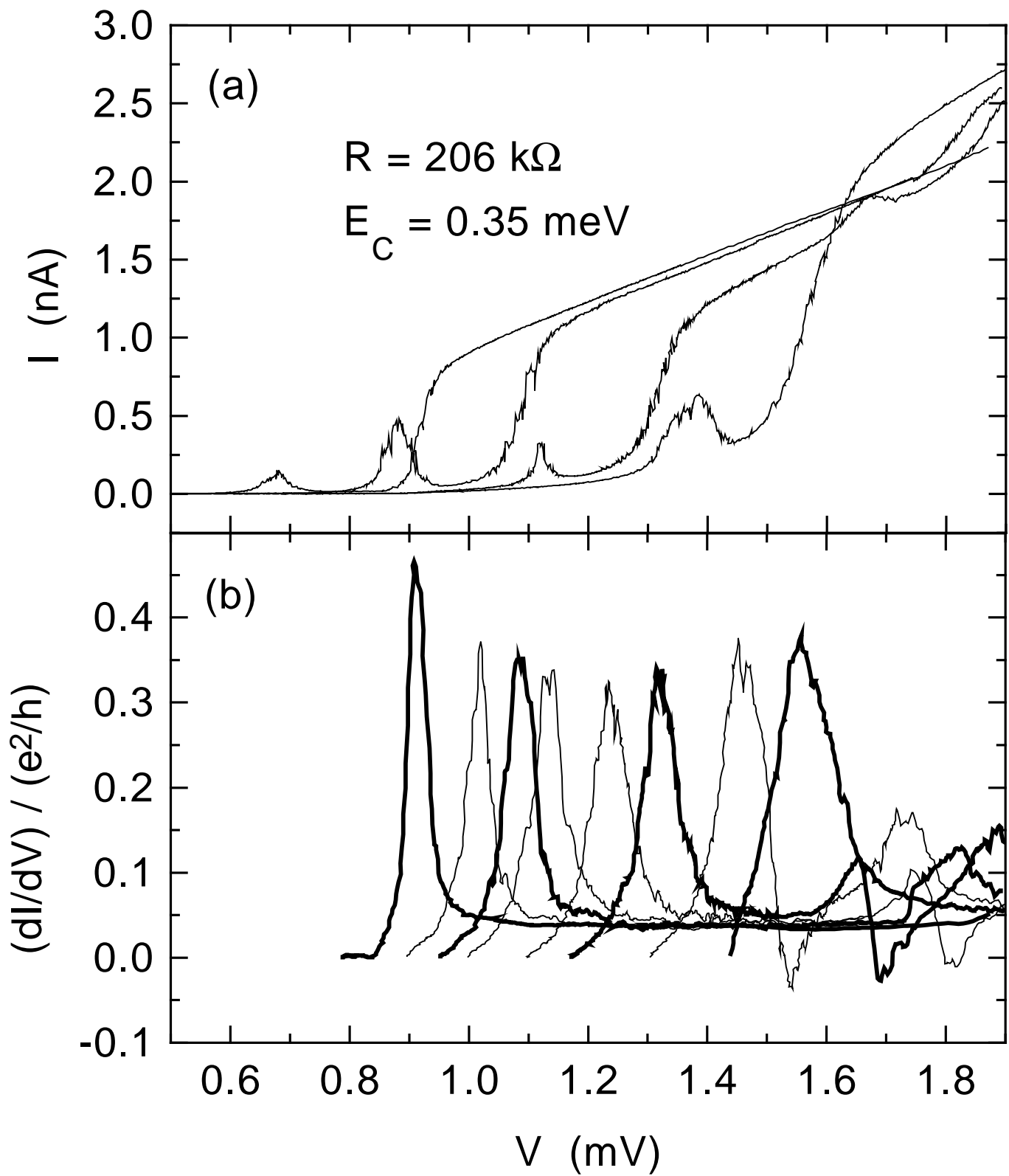
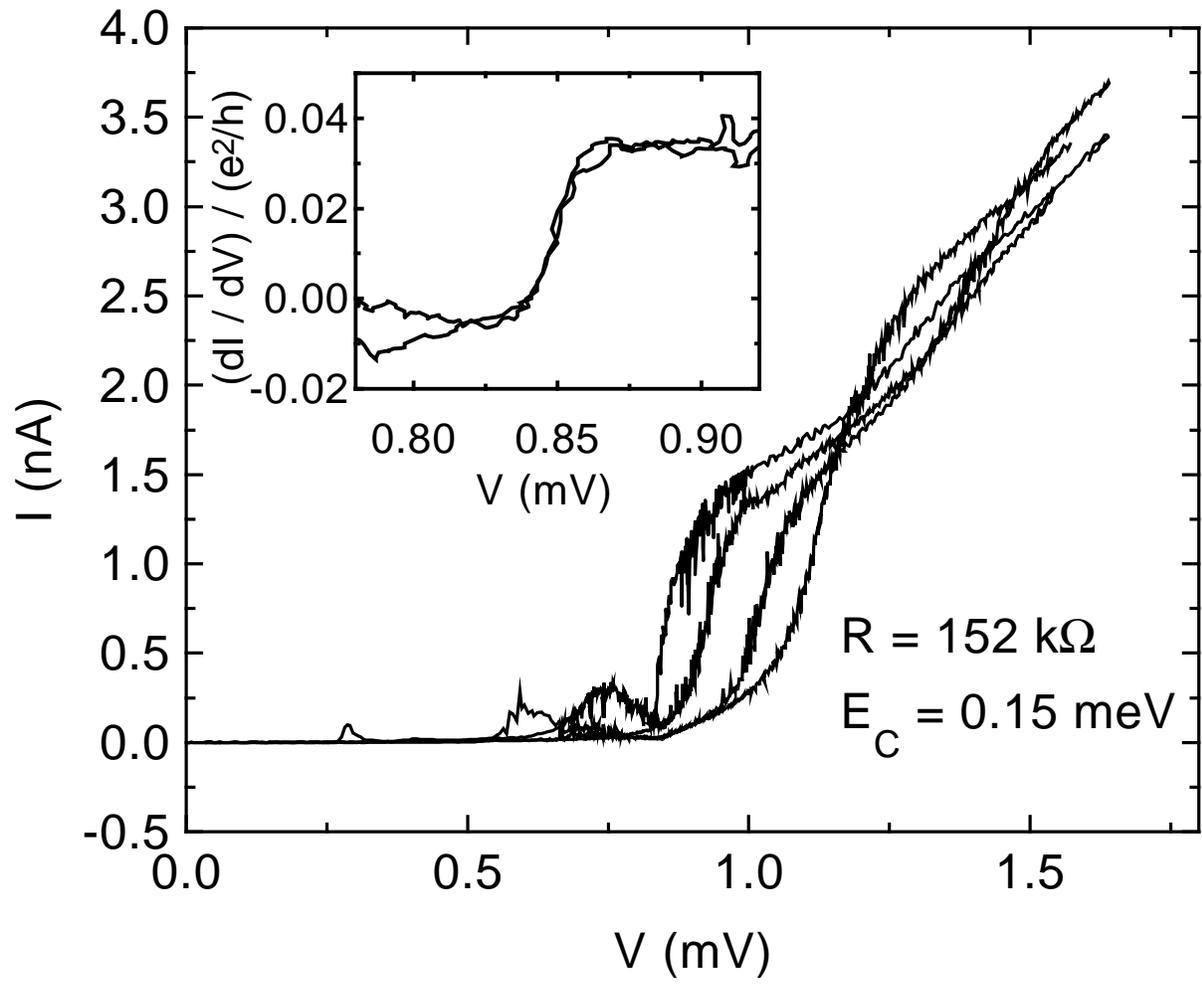
\end{document}